\documentclass[prb,twocolumn,superscriptaddress,amsmath,amssymb,letterpaper]{revtex4}
\usepackage{bm}

\def\beq{\begin{equation}}
\def\eeq{\end{equation}}
\def\beqn{\begin{eqnarray}}
\def\eeqn{\end{eqnarray}}

\begin{document}

\title{Multiferroic materials for spin-based logic devices}
\author{Rogerio de Sousa}
\altaffiliation{Current address: Department of
  Physics and Astronomy, University of Victoria, Victoria, BC V8W 3P6, Canada.}
\affiliation{Department of Physics, University of California,
Berkeley, CA 94720}
\author{Joel E. Moore}
\affiliation{Department of Physics, University of California,
Berkeley, CA 94720} \affiliation{Materials Sciences Division,
Lawrence Berkeley National Laboratory, Berkeley, CA 94720}
\date{\today}

\begin{abstract}
  Logical devices based on spin waves offer the potential to
  avoid dissipation mechanisms that limit devices based on
  either the charge or spin of mobile electrons.  Multiferroic
  magnetoelectrics, which are materials that combine ferroelectric
  and magnetic order, allow direct switching of magnetic order
  and thence of spin-wave properties using an applied electric
  field.  The intrinsic coupling between polarization and magnetic
  moments, generated by strong electronic correlations in these
  multiferroic materials, is argued to provide new approaches
  to spin-wave injection and spin-wave switching using
  applied voltages with no external magnetic field.  These
  effects are shown to arise in a phenomenological Landau
  theory of coupled electronic and magnetic orders in
  multiferroic BiFeO$_3$, and found to depend subtly on
  differences between the crystalline and film states of this material.
  \end{abstract}
%
%
\maketitle

\section{Introduction}

Some limitations of current semiconductor logic devices, such as their
relatively high power dissipation per operation, may be removed in
devices that use electron spin rather than electron charge to carry
out logic operations.  While ordered domains of electron spin are
currently the basic element of most information storage devices,
information manipulation (i.e., logic operations) is currently
accomplished using electron charge rather than spin.  Spin-based logic
is a relatively new field and, at the present time, has not
convincingly demonstrated advantages in practice over ordinary
charge-based logic.  In many cases, it seems that devices using spin
currents carried by {\it itinerant} electrons suffer from the same
dissipation mechanisms as ordinary electronic devices.

However, ferromagnetic and antiferromagnetic insulators provide
another route to spin currents: they support ``spin waves'' that can
transport spin and hence information over macroscopic distances,
without accompanying transport of charge.  This article discusses the
relevant physics of ``multiferroic magnetoelectric'' materials that
allow {\it direct electrical control} of the spin-wave propagation,
at room temperature and without external magnetic fields.  Recently
considerable effort has gone into demonstrating basic logic operations
with spin waves~\cite{kostylev05,khitun05}, as the dissipation mechanisms
in this approach to logic and the interactions between spin waves are
not well understood.  Multiferroic materials with both magnetic and
ferroelectric order address several of the most pressing problems in
making such spin-wave devices practical for applications: in the near
term, they allow electrical creation and detection of spin waves in a
compact volume and can hence serve as efficient converters between
electronic and spintronic logic.  In the longer term, multiferroic
materials may enable new designs for spintronic logic devices.

The value of multiferroic magnetoelectrics is that these strongly
correlated materials couple strongly to applied electric fields,
because of their ferroelectric character, while at the same time
showing magnetic order that enables propagation of spin waves.
Furthermore, the magnetic and ferroelectric orders are strongly
coupled, as in the recent demonstration of room-temperature switching
of magnetization direction by an electrical field in~\cite{zhao06}.
As an example of how this enables spin-wave manipulation, we will show
that the properties of a spin wave at wavevector ${\bf k}$, such as
its group velocity, depend on the angle between ${\bf k}$ and the
N\'{e}el magnetization direction.  The result of this combination of
effects is that electrical switching of the magnetization direction
modifies the spin-wave propagation sufficiently strongly to stop or
pass spin waves depending on the electrical state.  This functionality
is similar in concept to the function of ferroelectric materials as
electrooptic modulators~\cite{linesglass}.  While multiferroic
materials have a relatively long history~\cite{baryakhtar69,tilley82},
recent advances in fabrication and control of these materials have led
to a considerable resurgence of interest~\cite{spaldinreview}.

The theoretical calculations described in this paper are based
primarily on a time-dependent Ginzburg-Landau model for the free
energy of multiferroic materials.  The dynamical equations combine the
familiar Landau-Lifshitz dynamics of ferromagnetic and
antiferromagnetic materials, the standard macroscopic free energy for
polarization modes in ferroelectric materials, and additional terms
that couple the two orders.  The equations are solved to yield the
spectra of coupled polarization and spin-wave modes.  In choosing the
microscopic details of the multiferroic couplings, we focus for
simplicity on the material BiFeO$_3$.  In bulk crystals of BiFeO$_3$,
cycloidal antiferromagnetic order is established at 650 K, well below
the ferroelectric transition at 1120 K.  The next section introduces
the basic model for these crystals and explains one key result for
spin-wave manipulation in these crystals well below the transition
temperature: a zero-wavevector electric perturbation, such as an AC
electric field, couples to finite-wavevector spin waves, mediated by
the cycloidal magnetic structure in the ground state.  Films of
BiFeO$_3$ differ significantly from bulk crystals: they show uniform
antiferromagnetic order with a weak ferromagnetic component, rather
than cycloidal order.  Since the ability to create films is of
considerable importance for applications, Section III gives an overview
of how the model and results need to be modified for BiFeO$_3$ films,
and concentrates on how the switching process already demonstrated in
one experiment~\cite{zhao06} will modify spin wave propagation.  While
our focus will be on BiFeO$_3$ because it is the predominant
room-temperature multiferroic material, similar models can be applied
to 
other low-temperature multiferroic materials as well.

\section{Magnetoelectric coupling in bulk crystalline BiFeO$_3$: electronically active spin waves}

The first step in the theoretical description of this material is to
construct an appropriate phenomenological free energy.  At room
temperature, the system is far below its critical temperature for
antiferromagnetic and ferroelectric transitions, so that thermal
fluctuations can be ignored in the dynamics.  The long-period
incommensurate cycloid is produced in our model~\cite{desousa07} by a
Lifshitz term in the effective Ginzburg-Landau free energy; the
existence of such a term was previously
argued~\cite{sparavigna94,sosnowska82,zalesskii00,ruette04} as the
simplest symmetry-allowed explanation for the cycloidal order that is
observed in experiments unless a very large magnetic field is applied
(around 18 Tesla \cite{ruette04}).  The model free energy is
\begin{eqnarray}
F &=& {G L^4 \over 4} + {A L^2 \over 2} + {c' \sum_i
  (\nabla L_i)^2 \over 2} \cr
&& -\alpha \bm{P} \cdot \left[{\bf L} (\nabla \cdot {\bf L}) + {\bf L} \times
(\nabla \times {\bf L})\right] -\bm{P}\cdot \bm{E} \nonumber\\
&&+{r M^2\over 2}+ {a {P_z}^{2}\over 2}+{u{P_z}^{4}\over 4} + {a_\perp ({P_x}^2+{P_y}^2) \over 2} .
\label{fbulk}
\end{eqnarray} 
Here $L=|\bm{M}_1-\bm{M}_2|$ is a N\'{e}el vector describing the
staggered sublattice magnetization, $M=|\bm{M}_1+\bm{M}_2|$ is the
total magnetization of the material, and $P_z$ is the magnitude of the
ferroelectric polarization along ${\bf \hat z}$ [the cubic (111) and
equivalent directions in BiFeO$_3$]. This free energy is believed
to capture the most relevant physics of BiFeO$_3$ {\it bulk crystals},
while for films, the dominant coupling between magnetic and
ferroelectric orders is altered (see below).

If the Lifshitz term had been omitted, the AFM and FE orders would be
decoupled, leading to a simple ground state that is a commensurate,
isotropic Heisenberg antiferromagnet with $|{\bf L}_0|^{2}= {- A \over
  G}$, plus an easy-axis ferroelectric with polarization $\bm{P}_0=P_0
{\bf \hat{z}}$, $P_{0}^{2}= -a/u$.  Clearly $A < 0$ and $r>0$ for an
antiferromagnetic ground state, while $a<0$ and $a_{\perp}>0$ for a
ferroelectric ground state.  The Lifshitz term induces a cycloidal
ordering: the antiferromagnetic moment rotates in an arbitrary plane
including ${\bf \hat z}$.  It also increases the magnitude of $L_0$
and $P_0$.  For a reference ground state about which to consider small
perturbations~\cite{desousa07}, we take a cycloid with
\begin{equation} 
{\bf L}_0(x) = L_0 \left[\cos(q x) {\bf \hat z}
  + \sin(qx) {\bf \hat x}\right].  
\label{l0}
\end{equation} 
The cycloid direction is parallel to ${\bf \hat x}$ and the pitch is
$q = \alpha P_0 / c'$. The strength of antiferromagnetic and polar
orders are given by \beq L_{0}^{2} = {-A+\alpha^2 P_{0}^{2}/c' \over
  G},\quad P_{0}^{2}={-a+\alpha^2 L_{0}^{2}/c'\over u}.  \eeq

The linearized equations of motion for the
ferroelectric-antiferromagnet order parameters include the familiar
Landau-Lifshitz equations,
complemented by the Debye equation of motion for the ferroelectric
order parameter:
\begin{eqnarray}
\frac{\partial^{2} \bm{L}}{\partial t^{2}}
&=&-r(\gamma L_0)^2 \left[\frac{\delta F}{\delta \bm{L}}-\left(
{\bf \hat{L}}_{0}\cdot 
\frac{\delta F}{\delta \bm{L}}
\right)
{\bf\hat{L}}_{0}
\right],\label{ll}\\
\frac{\partial^{2} \bm{P}}{\partial t^{2}}&=&-f \frac{\delta F}{\delta \bm{P}}.\label{debye}
\end{eqnarray}
The constant $\gamma$ is a gyromagnetic ratio with dimensions of (sG)$^{-1}$,
while $f$ has dimensions of s$^{-2}$ and plays a similar role in the
ferroelectric equation of motion.

Details of the solution of these equations can be found elsewhere,
along with a description of their coupling to external electromagnetic
fields~\cite{desousa07}; in this work we would like to concentrate on
the consequences for the prospective applications discussed in the
Introduction.  The main result of interest for applications is that an
oscillating electric field at zero wavevector can couple to spin waves
at nonzero wavevector, as a consequence of the cycloidal order present
in the ground state.  At lowest order and with a perfectly harmonic
cycloid, the AC electric susceptibility has poles at the following shifted
phonon and antiferromagnetic frequencies:
\begin{eqnarray}
\Omega'^{2}_{{\rm PH}}&=& \frac{a_{\perp}}{\xi}+2c'q^2,\\
\Omega'^{2}_{{\rm AFM}}&=& 2c'q^2-\frac{2(\alpha q L_0)^2}{a_{\perp}},
\end{eqnarray}
where contributions of order ${\cal O} \left(\xi
  c'q^2/a_{\perp}\right)^{2}$ were dropped.  In these formulas, we have
defined $\xi = r(\gamma L_0)^2/f$ ($ \approx 10$ in BiFeO$_3$) and the
AC electric field is $\bm{E}\textrm{e}^{-i\omega t}$. The primed
frequencies are defined in units of $(\gamma L_0 \sqrt{r})\sim
10^{12}$~rad/s.

There are additional magneto-optical resonances that appear if a more
realistic theory is constructed by including, for example,
anharmonicity in the cycloidal ground state.  The possible impact of
these optically active spin waves for applications can be simply
stated: the cycloidal structure of the material means that even
zero-wavevector perturbations, which are potentially simpler to
generate than the stripline-launched waves in current
experiments~\cite{khitun05}, can couple to nonzero-wavevector spin waves
with a preferred direction.  Since antiferromagnetic spin waves
generically have large group velocity even at large wavelength, this
may provide a viable means for efficient launching of spin waves with
electrical fields.

The cycloidal nature of the magnetic order leads to yet another
potentially useful property for applications: The lowest frequency
branch for spin wave propagation is strongly anisotropic.  This occurs
due to the presence of a non-trivial symmetry operation, namely we may
rotate the cycloid pitch $\bm{q}$ out of the $xz$ plane with no energy
cost. This property implies the presence of a zero frequency mode with
dispersion independent of $k^{2}_{y}$:
\begin{equation}
\omega'^2\approx c'\left[k_{x}^{2}+k_{z}^{2}+\frac{3}{8}\frac{k_{y}^{4}}{q^2}
-\frac{k_{x}^{2}k_{y}^{2}}{q^2}\right],
\label{softmode}
\end{equation}
where we dropped contributions ${\cal O}(k^{6})$.  This soft mode
dispersion is strongly anisotropic, and may be useful for electrical
control of the spin wave group velocity via electrical switching of
the $\bm{P}_0$ direction. Note that the spin wave group velocity is
quite fast for propagation along the $xz$ plane, which is here defined
by the polarization $\bm{P}_0$ and the cycloid direction $\bm{q}$. For
this case the group velocity is given by $\gamma L_0 \sqrt{rc'}\sim
10^{6}$~cm/s. Interestingly, the anisotropic dispersion found in
Eq.~(\ref{softmode}) leads to zero group velocity (at $k\rightarrow
0$) for propagation along the $y$ direction (perpendicular to the
cycloid plane).

The next section discusses the surprisingly different physics that
occurs in BiFeO$_3$ films, where there is a uniform canted
antiferromagnetic ground state rather than a cycloid, in order to
predict how electrical switching of the magnetization direction will
modify the spin-wave propagation.

\section{Spin-wave switching in BiFeO$_3$ films}

The cycloid order is destroyed in epitaxially constraned BiFeO$_3$
films, as evidenced by the measurement of a much larger magnetic
susceptibility \cite{bai05} and confirmed by neutron scattering
experiments \cite{bea07}. For this reason, the Lifshitz invariant
assumed in Eq.~(\ref{fbulk}) can not be present or is too weak to
produce noticeable effects (for example, the cycloid period in films
is probably much larger than the typical domain size). Nevertheless,
the films seem to have a weak macroscopic magnetization, of the order
of $0.02 \mu_B$/Fe. This occurs even in the absence of applied fields
and can only be explained by canting of the sublattice magnetizations.
These considerations suggest the following model free energy for
BiFeO$_3$ films:
\begin{eqnarray}
F &=& \frac{a P_{z}^{2}}{2} + \frac{u P_{z}^{4}}{4}+
\frac{a_{\perp}(P_{x}^{2}+P_{y}^{2})}{2}-\bm{P}\cdot \bm{E}\nonumber\\
&&+ \sum_{j=1,2}\left[
\frac{r\bm{M}_{j}^{2}}{2}+\frac{G\bm{M}_{j}^{4}}{4}+
\frac{c' \sum_{i}\left(\nabla M_{ji}\right)^{2}}{2}\right]
\nonumber\\&&+J\bm{M}_{1}\cdot \bm{M}_{2}+ d\bm{P}
\cdot \bm{M}_{1}\times \bm{M}_{2}.
\label{ffilm}
\end{eqnarray}
Here $\bm{M}_{1}$ and $\bm{M}_{2}$ are sublattice magnetizations, and
$\bm{P}$ is the ferroelectric polarization. We assumed~\cite{fox77},
that canting occurs due to a linear magnetoelectric effect through a
Dzyaloshinskii-Moriya vector equal to $d\bm{P}$.  As a result, this
free energy is invariant under inversion at a center located in
between the two sublattices.  This coupling is expected if the
low-temperature phase is imagined as resulting from symmetry breaking
in a high-temperature cubic phase, but another possibility has been
suggested based on electronic structure calculations~\cite{ederer05}.
That work finds that the dominant contribution to the
Dzyaloshinskii-Moriya vector is independent of $\bm{P}$.  In this
case, the ferromagnetic vector $\bm{M}$ is nearly decoupled from
$\bm{P}$.  Experimental tests of how polarization switching modifies
the magnetic structure in films are a high priority; tests so far
indicate that the antiferromagnetic vector $\bm{L}$ switches to remain
perpendicular to $\bm{P}$, but the response of $\bm{M}$
has not been probed yet.

The reference ground state for film structures is an uniform-canted state,
\begin{subequations}
\begin{eqnarray}
\bm{M}_{01}&=&M_0 \left(\sin{\beta} \hat{\bm{x}} +\cos{\beta}\hat{\bm{y}}\right),\\
\bm{M}_{02}&=&-M_0 \left(-\sin{\beta} \hat{\bm{x}} +\cos{\beta}\hat{\bm{y}}\right),
\end{eqnarray}
\end{subequations}
with canting angle approximately given by $\tan{\beta}\approx
dP_0/(2J)$.  The spin waves for this canted state were calculated a
long time ago for propagation along the symmetry axis ($\bm{P}_0$ in
our case), perpendicular to the canting plane \cite{herrmann63}. There
are two different spin wave modes: One is a low frequency mode, with
linear dispersion $\tilde{\omega}=\sqrt{2Jc'}k$, where
here $\tilde{\omega}=\omega/(\gamma M_0)$. The other is a high frequency
mode, with gap equal to the Dzyaloshinskii-Moriya coupling,
$\tilde{\omega}\approx dP_0$. 

We now discuss the possibility of electronic excitation of spin waves
in a canted multiferroic such as BiFeO$_3$ films.  The canted-uniform
magnetic order of the films imply that an electrical probe with
wavevector $k$ will only excite spin waves at the same wavector $k$.
This is in drastic contrast to the case of bulk BiFeO$_3$ described
above, where the cycloidal magnetic order allowed the transduction of
a $k=0$ excitation into an $nq$ excitation. 

However, the presence of a macroscopic weak magnetization in the film
state allows a whole new set of magnetoelectric functionalities.
Below we provide a summary of our results, with a focus on electrical
control of magnon propagation. A detailed description is available
elsewhere \cite{desousa07b}. For example,
the fact that the weak magnetization scales linearly with the electric
polarization ($2M_0 \sin{\beta}\approx dP_0 M_0 /J$) allows electrical
excitation of low frequency spin waves provided the AC electric field
is polarized along the weak magnetization direction ($x$ direction in
our case). Moreover, an electric field polarized in the $yz$ plane
(perpendicular to the weak magnetization) will excite a high frequency
magnon. The latter effect may be observed optically in the far
infrared region, since this will lead to an electromagnon resonance (a
pole in the dielectric function) at the Dzyaloshinskii-Moriya
frequency. 

Moreover, the presence of a macroscopic weak magnetization will also
lead to a sizable anisotropy in the low frequency magnon dispersion.
The physical origin of this anisotropy is completely different than
the one occurring in bulk BiFeO$_3$ [See Eq.~(\ref{softmode}) above].
In films, the presence of a weak magnetic moment implies the spin waves
will produce a macroscopic AC magnetization with finite amplitude $\delta
\bm{M}$. From Maxwell's equations we see that this must create a
magnetostatic field of the form
\begin{equation}
\bm{h}=-4\pi \left(\delta\bm{M}\cdot \hat{\bm{n}}\right) \hat{\bm{n}}\textrm{e}^{i(\bm{k}\cdot \bm{r}-\omega t)},
\label{hind}
\end{equation}
where $\hat{\bm{n}}$ is the spin wave propagation direction,
$\bm{k}=k\hat{\bm{n}}$. The magnetostatic field contributes a term $2\pi
(\delta \bm{M}\cdot \hat{\bm{n}})^2$ to the free energy, tending to
increase the spin wave frequencies whenever the quantity
$\delta\bm{M}$ has a finite projection along $\hat{\bm{n}}$. As a
consequence, the low frequency mode becomes anisotropic according to
\begin{eqnarray}
\tilde{\omega}^{2}(\bm{k})&\approx& 
2J c'\left( 1+\frac{4\pi}{J}n_{z}^{2}\right)  k^{2} + 
\frac{4\pi (dP_0)^{2}}{J}n_{y}^{2}.
\label{softfilm}
\end{eqnarray}
Interestingly, a gap of the order of $\sqrt{4\pi/J}(dP_0)$ appears for
spin waves propagating along the $y$ direction (the direction of the
N\'{e}el vector in our case). Similarly to the case of bulk BiFeO$_3$,
this effect may be used as an electrical switch for spin waves
propagating along the $xz$ direction. Changing the orientation of
$P_0$ electrically will drastically reduce the group velocity of a
spin wavepacket propagating along the $xz$ plane.

The physical origin of this gap lies in the anticrossing between the
photon dispersion ($\omega=ck$) and the low frequency magnon
dispersion, a phenomenon analogous to the formation of polaritons for
optical phonon excitations. The magnetostatic propagation anisotropy
Eq.~(\ref{softfilm}) arises precisely because the spin waves travel
with finite $k>\omega/c$, which is always greater than a few cm$^{-1}$
due to the small domain sizes. For this reason, the magnon dispersion
is effectively gapped, but we emphasize that the strict $k\rightarrow
0$ limit, observable only at very long wavelengths (several
centimeters) has $\omega\rightarrow 0$ and no orientation dependence
as expected by symmetry.

\section{Conclusions}

In summary, we discussed the possibility of electrical excitation and
control of spin waves in BiFeO$_3$ monocrystals and heterostructures.
The presence of the inhomogeneous cycloidal order in bulk BiFeO$_3$
implies an AC electric field at $k\approx 0$ will excite spin waves at
integer multiples of the cycloid wavevector (at $k=nq$). This shows
that bulk BiFeO$_3$ is potentially useful as a source for finite
wavevector magnons, which may become a viable alternative to the
current strip-line designs \cite{covington02}. Although BiFeO$_3$
films do not have this functionality, spin waves may still be excited
electrically, due to the presence of a Dzyaloshinskii-Moriya linear
magnetoelectric effect. 

We remark that changing the direction of the ferroelectric
polarization may force a switch of the N\'{e}el vector orientation in
a special class of multiferroic materials, an effect already
demonstrated in BiFeO$_3$ films \cite{zhao06}. In our work we showed
in addition that BiFeO$_3$ monocrystals as well as heterostructures
possess a sizable anisotropy in the lowest frequency spin wave
dispersion.  The combination of these two physical phenomena implies
that the magnon group velocity may be stopped electrically in
BiFeO$_3$, an important new functionality for devices based on spin
wave propagation. Similar effects may be demonstrated in other well
studied multiferroic materials, such as TbMnO$_3$ \cite{kimura03}, or
BaNiF$_4$ \cite{ederer06}.

The authors acknowledge useful conversations with J. Orenstein and R.
Ramesh.  This work was supported by WIN (RdS) and by NSF DMR-0238760
(JEM).

\end{document}